# Fatigue-Related Reaction Time Forecasting via EEG Functional Connectivity in Sustained Attention Task

Bo Sun[1,2], Liang Ma[1,2]

[1] Department of Industrial Engineering, Tsinghua University, Beijing 100084, China
[2] National Key Laboratory of Human Factors Engineering, Tsinghua University, Beijing 100084, China

***Abstract*—Mental fatigue related behavioral performance decline precipitates catastrophic accidents in sustained attention tasks. While existing neurophysiological systems effectively detect current behavioral performance, they often lack the capability to forecast behavioral lapses with sufficient temporal lead time for intervention. This study proposes a novel model for the reaction time (RT) forecasting using EEG functional connectivity features. Thirty participants engaged in a sustained Psychomotor Vigilance Test (PVT) with concurrent 30-channel EEG recording. Mutual information (MI) between electrodes was calculated as functional connectivity features. Random Forest regression model (RF) was trained to predict single-trial RTs across forecasting horizons ranging from 0 to 20 seconds. The model demonstrated robust predictive validity, achieving a Root Mean Square Error (RMSE) of 23.75 ms for immediate detection and maintaining high accuracy (RMSE = 24.07 ms) across different forecasting horizons. Interpretability analysis via SHAP and Linear Mixed Effects model further support the validity of the proposed model and revealed distinct temporal biomarkers. This study validates the feasibility of forecasting behavioral performance 20 seconds in advance, offering a promising methodology for proactive fatigue management in safety-critical systems.**

## I. Introduction

Sustained attention tasks are pervasive in safety-critical domains such as transportation, healthcare, and industrial operations, where prolonged wakefulness and high cognitive demands are commonplace [1]. In these environments, mental fatigue manifests as significant behavioral instability, characterized by delayed reaction times and attentional lapses [2] that can precipitate catastrophic accidents [3], and the rapid behavioral impairment is inevitable [1]. Consequently, the ability to predict behavioral impairment in advance is of paramount importance for proactive accident prevention [4]. In high-intensity working environments, fatigue forecasting holds significant translational value by enabling timely interventions. It allows personnel to deploy necessary countermeasures before performance critically declines, thereby averting irreversible operational errors [5]. Furthermore, in non-interruptible critical tasks, anticipating potential behavioral impairments facilitates the scientific planning of subsequent task loads [6].

To address the critical need for predicting behavioral impairment, significant research efforts have been directed toward predicting reaction time (RT) as a continuous, objective index of cognitive state. Electroencephalography (EEG) has emerged as the preferred modality [7] due to its ability to offering the temporal resolution necessary to track instantaneous fluctuations in sustained attention. Various computational models have been put forward to predict reaction time with EEG time-frequency features in diverse sustained attention tasks, including simulated flight [8], autonomous driving [4], psychomotor vigilance test [9], etc.

Apart from the perspective of independent contribution of each brain region in fatigue related performance, an integrated perspective that stress the fundamental network architecture of the brain has emerged in recent year.

According to the Global Neuronal Workspace Theory [10], effortful cognitive processes such as sustained attention are physiologically represented by the integrated activity of a large ensemble of "workspace neurons" distributed across the cortex. As highlighted by Ishii et al. [11], fatigue mechanisms involve a complex interplay between mental facilitation and inhibition systems that modulate task-related regions to regulate performance. Consistent with these theories, convergent evidence suggests that mental fatigue is not merely the result of impaired local activity, but rather involves a systemic reorganization of functional connectivity among brain regions [7]. Consequently, recent studies have increasingly incorporated network-level features in reaction time prediction, including Riemannian geometry features [12], differential phase synchrony representations [13], phase-locking value and covariance representations [14], [15], etc. Collectively these previous models achieved high predictive accuracy, advancing the prediction of reaction time in sustained attention tasks.

Despite the high prediction accuracy achieved by prior studies, limited research has focused on the forecasting of reaction time. From the perspective of automotive safety, forecasting declining arousal is critical for issuing timely warnings and mitigating accident risks [16]. A warning provided even 20 seconds before the performance offers operators and systems sufficient latency to adopt micro intervention [16], [17], such as bright light stimulation [18], auditory alerts [19], brief sensory stimulation [20] or just pause ongoing work, to mitigate imminent risks. Consequently, the practical utility of such a predictive system is fundamentally determined by its forecasting horizon, the duration by which the prediction precedes actual behavioral performance. However, the majority of the previous models rely exclusively on EEG features extracted adjacent to the behavioral response[8], [10], [13], [14], [16], hindering their utility to make forecasts, thus impedes the possible remedies to behavioral impairment. Stikic et al. [21] successfully forecasted reaction time 15 minutes in advance during sustained attention tasks with EEG features, but their methodology relied on 5-minute averaging windows. This extended averaging process obscures the rapid, moment-to-moment fluctuations characteristic of dynamic behavioral changes, ultimately hindering its applicability for real-time intervention. Similarly, Ayaz et al. [22] explored the time-lagged correlation between EEG feature and reaction time, demonstrating the relationship between the past neural activities and the future performance in a more dynamic approach. However, their work stopped short of translating this relationship into a precise, quantitative prediction model.

Consequently, there remains a vital need to develop a predictive model that utilize short feature extraction windows to forecast moment-to-moment behavioral performance in sustained attention tasks, enabling the detection of impending behavioral impairments prior to their manifestation.

To address this critical gap, the current study employed a Psychomotor Vigilance Test (PVT) with concurrent EEG recording to capture the neural activities during the experiment. A random forest regression model based on functional connectivity features was subsequently utilized to generate forecasts of reaction time. Furthermore, by integrating this regression model with SHapley Additive exPlanations (SHAP) [23], our study sought to elucidate the contributions of specific network connections.

## II. METHOD

### A. Participants

Thirty healthy volunteers were recruited for this study (12 females, mean age = 23.07 ± 3.58. All participants were right-handed and possessed normal or corrected-to-normal vision. Inclusion criteria required participants to be free of fatigue disorders, maintain over 7 hours of sleep for the two preceding nights, and abstain from stimulants prior to the experiment. The experimental procedures were conducted in strict accordance with the Declaration of Helsinki and were approved by the local ethics committee at the Department of Industrial Engineering, Tsinghua University (THU-04-2026-0027). All participants provided written informed consent prior to the experiment and received monetary compensation for their participation.

### B. Experiment Procedure

Upon arrival, participants signed informed consent and were prepared for EEG data acquisition. The experiment was conducted in a sound-attenuated, dimly lit room, where participants were seated approximately 60 cm in front of a 22-inch LCD monitor (DELL, USA). To assess changes in subjective mental states, participants completed two questionnaires immediately before and after the experimental task, including the Chalder Fatigue Scale (CFS, [24]) and the Stanford Sleepiness Scale (SSS, [25]).

Following the pre-assessment, participants went through a Psychomotor Vigilance Test (PVT) to maintain sustained attention [26]. Stimuli were presented using PsychoPy [27]. During the test, participants were instructed to monitor a computer screen and press space button as quickly as possible upon the appearance of a millisecond counter. The counter appeared at the center of the screen with a random inter-stimulus interval (ISI) ranging uniformly from 2 to 10 seconds. The stimulus remained active until a response was made. Upon a valid response, the reaction time was displayed for 1 second to provide immediate performance feedback. If the stimulus remained on the screen for 2 seconds without a response, the program displayed a “time-out” notification and proceeded to the next trial. The formal experiment consisted of 400 trials and lasted approximately 40 minutes. The procedure is shown in Fig. 1.

### C. EEG Data Acquisition and Preprocessing

EEG signals were recorded using a wireless 30-channel system (NeuSen.W, Neuracle, China) with a sampling rate of 1000 Hz. Electrodes were positioned according to the international 10-20 system at the following sites: Fp1/2, Fz, F3/4, F7/8, FC1/2/5/6, Cz, C3/4, T7/8, CP1/2/5/6, Pz, P3/4/7/8, PO3/4, Oz, O1/2. Throughout the experiment, electrode impedances were maintained below 10 kΩ.

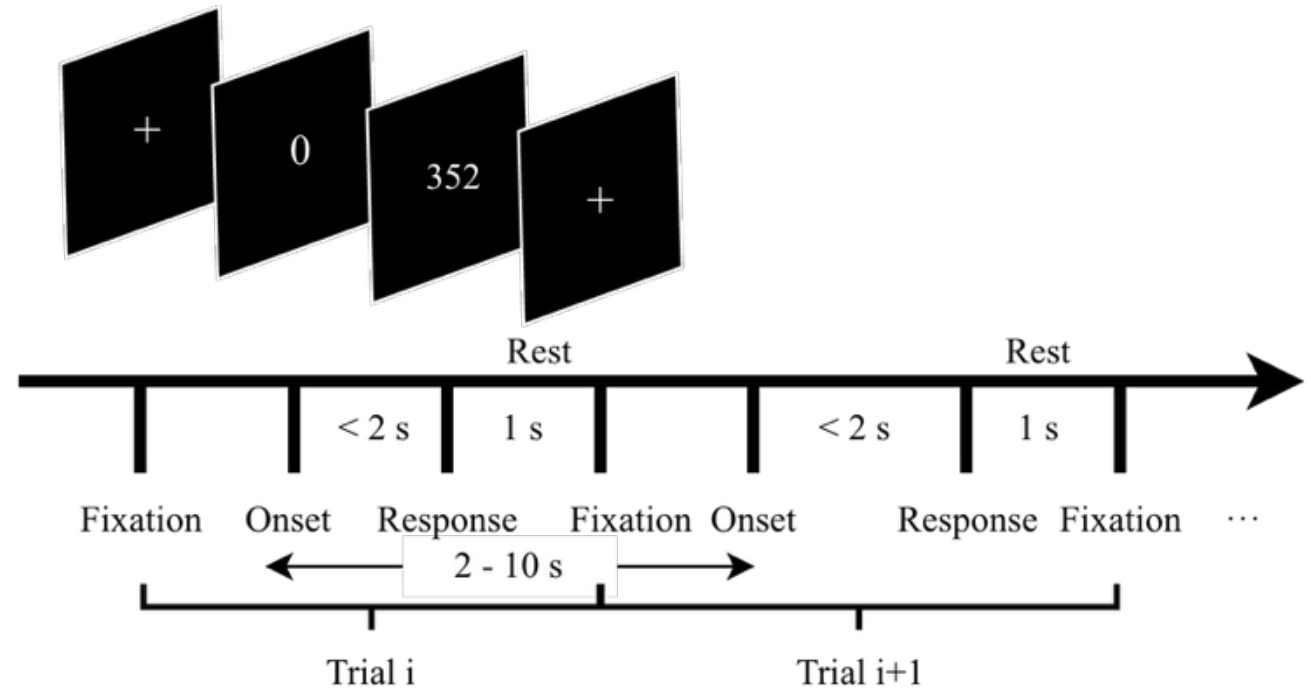

Fig. 1. **Experimental timeline of a single Psychomotor Vigilance Test (PVT) trial.** Participants wait for a random interval (2–10 s) before a millisecond counter appears. Responses within the 2-second window yield 1 second of reaction time feedback, whereas an omission results in a "time-out" notification.

Data preprocessing was performed in MATLAB (Mathworks 2025a, Natick, MA) using the EEGLAB toolbox [28] and the DISCOVER pipeline [29]. Specifically, continuous data were first bandpass filtered between 0.1 and 47 Hz using a zero-phase Hamming-windowed sinc FIR filter (order = 33,000) and subsequently downsampled to 500 Hz. Bad channels were identified and removed based on three criteria: (1) a flatline duration exceeding 5 seconds; (2) a low correlation with neighboring channels (i.e., the channel could not be predicted by a random subset of other channels for at least 80% of the recording); or (3) a z-scored noise-to-signal ratio greater than 4 [30]. The remaining signals were re-referenced to the common average reference. Artifacts were removed using automated Independent Component Analysis (ICA). Components classified as 'Muscle' or 'Eye' with a probability greater than 80% were subtracted from the data [30]. Finally, removed channels were interpolated using spherical splines [31].

### D. Feature Extraction

Functional connectivity was employed as the primary feature set for fatigue prediction. Specifically, we calculated the mutual information between all possible pairs of EEG channels. MI was selected as the connectivity metric based on several reasons. First, MI provides a comprehensive measure of statistical dependence capable of capturing both linear and non-linear interactions [32], which fit the nature of EEG activities as neural dynamics are inherently non-linear and non-stationary, particularly during transitions between cognitive states [33]. Moreover, MI does not rely on assumptions of Gaussian distribution, which offers a robust estimation of coupling in complex biological signals. Second, MI demonstrates robustness against the effects of volume conduction when calculating sensor-level connectivity [34], thereby reducing the risk of spurious correlations arising from signal leakage. Third, the low computational cost of MI facilitates the processing of high-dimensional feature spaces, making it particularly well-suited for the sliding-window calculations required for real-time, moment-to-moment predictions. For two discrete time series $X$ and $Y$, the mutual information is calculated as:

$$I(X;Y) = \sum_{y \in Y} \sum_{x \in X} p(x,y) \log\left(\frac{p(x,y)}{p(x)p(y)}\right) \quad (1)$$

where $p(x)$ and $p(y)$ denote the marginal probability distributions of $X$ and $Y$ respectively, and $p(x,y)$ represents their joint probability distribution.

Feature extraction was performed using a 5-second window follow previous research [12], [13], [14]. To ensure the stability of the connectivity estimates, this 5-second window was segmented into consecutive non-overlapping 1-second epochs. The MI was computed for each 1-second epoch and subsequently averaged for each channel pair per window.

To investigate the feasibility of fatigue forecasting, we extracted features at varying temporal distances from the event onset. We defined 20 distinct horizons, where the lag between the end of the analysis window and the onset of the reaction stimulus ranged from 0 to 20 seconds (in 1-second increments). To ensure data completeness and signal stationarity, feature extraction was restricted to trials starting from the 9th trial of the PVT task.

### E. Model Construction

Data from each participant were individually segmented into training and testing subsets and were further pooled to construct an aggregate training and testing dataset, in order to ensure the balanced representations of data from different participants.

The predictors consisted of 435 functional connectivity features (30x29/2) extracted from a specific time lag. The target variable was the corresponding Reaction Time (RT). Trials with RT shorter than 100 ms were classified as false alarms, whereas trials with RT longer than 500 ms were classified as lapses. Both false alarms and lapses were excluded from the analysis to focus on valid response variability [35]. To reduce trial-to-trial noise, the RTs for each participant were then smoothed using a moving average window with length of 5 trials.

In our study, a traditional machine learning approach was deliberately chosen over deep learning alternatives to prioritize model interpretability.

Random forest regression model (RF) was applied in the study. RF is an ensemble learning method that operates by constructing a multitude of decision trees during training and outputting the average prediction of the individual trees [36]. This model demonstrates high robustness against overfitting through bootstrap aggregation (bagging), possesses inherent insensitivity to feature scaling, and is capable of capturing the complex, non-linear relationships characteristic of neural data without requiring extensive parameter tuning. Previously, Peng et al. [37] applied RF to predict mental fatigue levels using fNIRS-based functional connectivity, supporting its validity in network-based RT prediction.

Hyperparameters of RF were optimized using Optuna [38], an automatic hyperparameter optimization software framework designed for efficient and robust search strategies. The framework utilized a define-by-run principle and Tree-structured Parzen Estimator (TPE) as a sampler. The objective function maximizes the the Out-of-Bag (OOB) score, and the number of trials is 50. The utilized hyperparameters of RF and their searching spaces are presented in Table I.

Table I Hyperparameter searching space for Optuna

| Hyperparameter | Searching space |
|---|---|
| Max features | 'sqrt', 'log2', 0.3, 0.5, 0.8 |

| Minimum samples per lead | 1-50 |
| --- | --- |
| Maximum depth | 10-50 |

Hyperparameter tuning was performed on the detection model (using features extracted from the 0-second lag). The optimization yielded the following configuration: max_features = 0.3, min_samples_leaf = 4, and max_depth = 48. Combining n_estimators = 200, these optimal hyperparameters were subsequently applied to all models across the varying forecasting horizons (1 s to 20 s lags).

### *F. Model Validation and Interpretability*

Model performance was evaluated using 10-fold cross-validation. Performance was quantified using the Root Mean Square Error (RMSE) and Pearson's correlation coefficient between the predicted and actual RTs.

To calculate the contribution of specific functional connectivity features, we employed SHapley Additive exPlanations (SHAP), a game-theoretic approach to interpretable machine learning that assigns each feature an importance value representing its contribution to the prediction [23]. The final models of each horizon were refitted to the entire dataset. SHAP values were calculated for every feature in each model to identify the neural connectivity patterns most predictive of reaction times.

### *G. Statistical Analysis*

To verify the successful induction of fatigue, paired samples t-tests were conducted to compare behavioral measures obtained at the beginning and end of the experiment. Specifically, we compared subjective ratings (SSS and CFS) collected pre- and post-task, as well as mean reaction times (RT) derived from the first 20 and last 20 trials. Additionally, to validate the association between subjective fatigue levels and behavioral performance, Pearson's correlation coefficients were calculated between post-task subjective ratings and the mean RT of the final 20 trials.

To elucidate the specific contributions of the predictive features, the analysis proceeded in two stages. At the channel level, we first identified the top-five features with the highest mean absolute SHAP values. The directionality of their impact on reaction time was assessed by computing Pearson's correlation coefficients between the original connectivity strength of each feature and its corresponding SHAP value. To obtain a more integrated view, functional connectivity features were grouped into regional connections (e.g., frontal-frontal, frontal-occipital) based on the channel location: frontal, parietal, occipital, left temporal, and right temporal. The detailed correspondence between channels and regions is presented in Appendix A). Subsequently, a Linear Mixed Effects (LME) model was applied to the SHAP values derived from the RF to evaluate the contribution of each regional connection. The LME model was specified with the SHAP value as the dependent variable and the regional connection type as a fixed effect, while including random intercepts for both channels involved in the connection to account for dependencies arising from features sharing common electrodes [SHAP value ~ 1 + Region Pair + (1|Chan. 1) + (1|Chan. 2)]. Then, the significance of each region pair's contribution was evaluated. False Discovery Rate (FDR) correction was conducted for the multiple comparisons of different region pair.

## III. RESULT

### *A. Behavioral Performance*

Analyses of subjective ratings and reaction time confirmed the successful induction of mental fatigue. Post-task ratings on the Stanford Sleepiness Scale (SSS) ($t(29) = 10.45$, $p < .001$, Cohen's $d = 1.91$) and Chalder Fatigue Scale (CFS) ($t(29) = 9.90$, $p < .001$, Cohen's $d = 1.81$) were significantly higher than pre-task ratings (Fig. 2a, b). Paralleling this increase, behavioral performance showed a marked decline. Mean reaction times (RTs) during the final 20 trials were significantly larger compared to the first 20 trials ($t(29) = 10.85$, $p < .001$, Cohen's $d = 1.98$; Fig. 2c). Furthermore, Pearson's correlation coefficients revealed that the prolonged RTs at the end of the task were positively associated with post-task SSS scores ($r = 0.52$, $p = .003$) and CFS scores ($r = 0.38$, $p = .037$; Fig. 2d, e), corroborating the link between subjective fatigue and behavioral performance.

### *B. Model Performance*

The detection model (0-s lag) demonstrated a strong correspondence between predicted and actual reaction times, achieving a Pearson's correlation coefficient (r) of $0.691 \pm 0.014$ and a Root Mean Square Error (RMSE) of $23.752 \pm 0.645$ ms across 10-fold cross-validation.

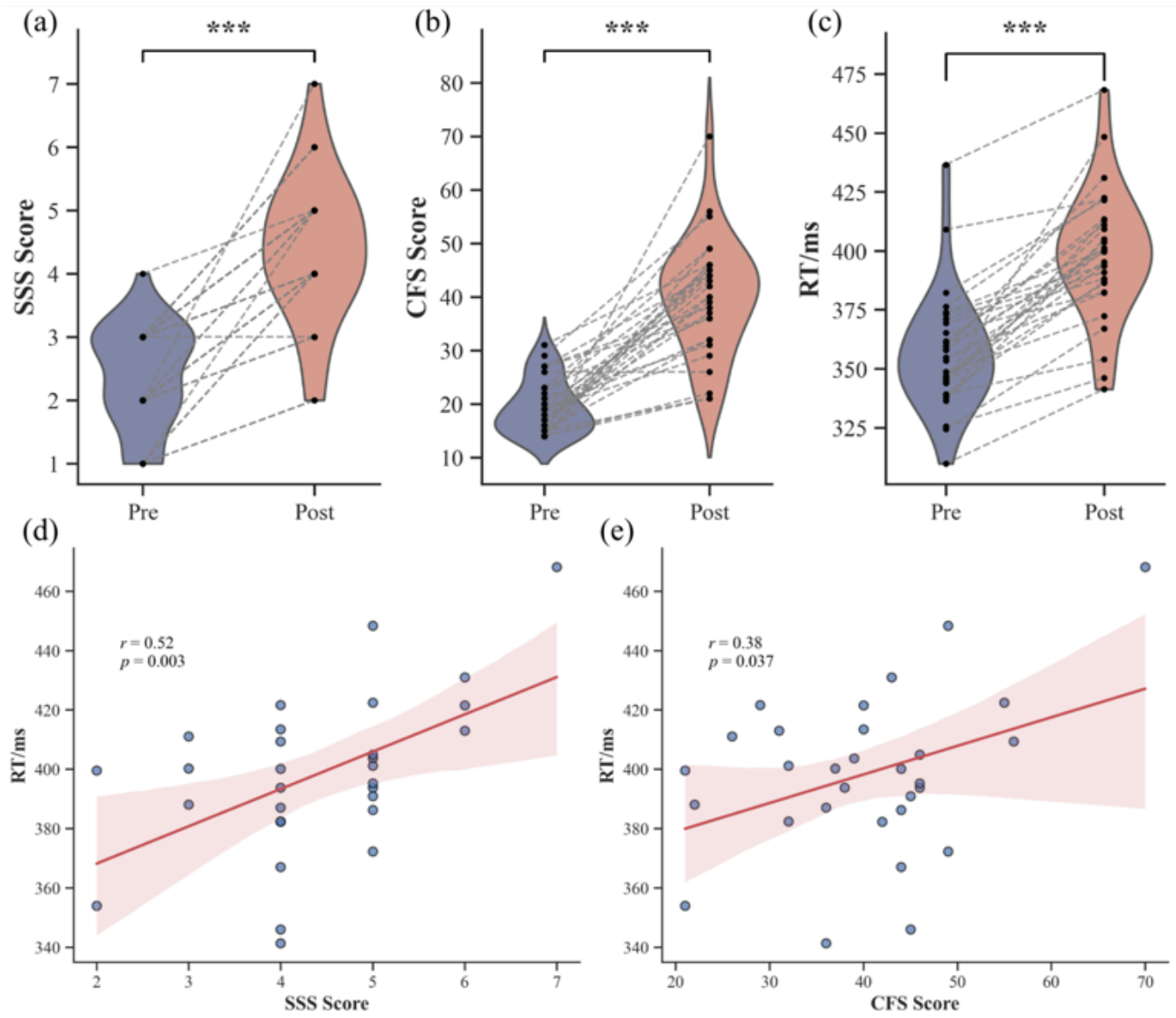

Fig. 2. **Behavioral Performance.** Violin plots illustrate significant increases from pre- to post-task in (a) Stanford Sleepiness Scale (SSS) scores, (b) Chalder Fatigue Scale (CFS) scores, and (c) mean reaction times (RT). Scatter plots show positive correlations between post-task RTs and (d) SSS scores, as well as (e) CFS scores.

Across the forecasting models with varying time lags, performance was poorest at the 15-s lag, yielding the highest RMSE (24.187 ± 0.570 ms) and the lowest correlation (r = 0.678 ± 0.012). At the maximum lag of 20 s, the model maintained high predictive accuracy, with a Pearson's correlation coefficient of 0.680 ± 0.013 and an RMSE of 24.066 ± 0.554 ms. Detailed cross-validation performance metrics for each model are presented in Table II. These results demonstrate that the proposed model based on functional connectivity features can effectively predict reaction times in advance, confirming the feasibility of fatigue forecasting.

### C. Model Parameter Weight Analysis

SHAP value analysis revealed distinct contribution of specific channel-level connections. Four connections consistently emerged as top-five predictors across nearly all horizons. Specifically, the connections between channel O1 and P7 (r = 0.87 - 0.91) and the connections between channel PO4 and P8 (r = 0.79 - 0.87) displayed a robust positive influence on reaction time. Similarly, the connection between channel F3 and F4 ranked among the top-five features across lags 1–20, also exhibiting a consistent positive association (r = 0.71 - 0.78). In contrast, the connection between channel O1 and Oz demonstrated a robust negative influence on reaction time (r ranged from -0.86 to -0.81). Several other connections exhibited predictive significance within specific models. The connection between channel PO4 and Fp2 appeared as top-five feature across diverse intervals (lags 4 – 8, 11 – 17 and 19 s), showing a positive influence (r = 0.72 - 0.81). The connection between channel T7 and F7 was prominent in lags 0, 9 and 10 s, displaying a negative influence (r = -0.72 to -0.67). Additional sporadic positive predictors included P4-CP6 connection (lags 1, 2 s, r = 0.74, 0.77), PO3-F3 connection (lags 0 s, r = 0.76), PO3-F4 connection (lags 18, 20 s, r = 0.71, 0.78), and PO4-F3 connection (lags 3 s, r = 0.64). The directionality and relative importance of these features are further illustrated in Fig. 3, which provides representative SHAP beeswarm plots and topoplots for detection and forecasting models. Appendix B demonstrates top 5 features in all models.

Table II Performance metrics of each model

| Lag/s | RMSE | r |
|---|---|---|
| 0 | 23.752 ± 0.645 | 0.691 ± 0.014 |
| 1 | 23.869 ± 0.615 | 0.686 ± 0.013 |
| 2 | 23.949 ± 0.605 | 0.684 ± 0.014 |
| 3 | 24.083 ± 0.615 | 0.679 ± 0.011 |
| 4 | 24.065 ± 0.600 | 0.680 ± 0.011 |
| 5 | 24.072 ± 0.707 | 0.680 ± 0.013 |
| 6 | 24.024 ± 0.656 | 0.681 ± 0.012 |
| 7 | 23.960 ± 0.679 | 0.684 ± 0.013 |
| 8 | 23.945 ± 0.624 | 0.684 ± 0.013 |
| 9 | 23.963 ± 0.670 | 0.684 ± 0.014 |
| 10 | 23.962 ± 0.644 | 0.683 ± 0.015 |
| 11 | 24.015 ± 0.631 | 0.682 ± 0.015 |
| 12 | 24.065 ± 0.627 | 0.680 ± 0.015 |
| 13 | 24.046 ± 0.545 | 0.681 ± 0.014 |
| 14 | 24.010 ± 0.551 | 0.682 ± 0.013 |
| 15 | 24.129 ± 0.567 | 0.678 ± 0.012 |
| 16 | 24.187 ± 0.570 | 0.676 ± 0.013 |
| 17 | 24.014 ± 0.574 | 0.682 ± 0.014 |
| 18 | 24.065 ± 0.588 | 0.680 ± 0.013 |
| 19 | 24.024 ± 0.560 | 0.682 ± 0.012 |
| 20 | 24.066 ± 0.554 | 0.680 ± 0.013 |

LME model further revealed significant contributions of regional connections in different models. Two regional connections exhibited consistent predictive significance across all time lags, including the connection between the left temporal and occipital regions ($\beta$ = 0.24 - 0.30, all ps < .05) and the connection within the occipital region ($\beta$ = 0.25 - 0.36, all ps < .05). Three regional connections emerged as significant predictors in a subset of the models. Specifically, the connection within the frontal region displayed significant effect in model with lag 0, 7 - 9, 11, 13 – 20 s ($\beta$ = 0.12 - 0.16, ps < .05), the connection between frontal and left temporal regions displayed significant effect in model with lag 19 s ($\beta$ = 0.10, p = .049), the connection between frontal and occipital regions displayed significant effect in model with lag 14 – 20 s ($\beta$ = 0.11 - 0.13, ps < .05), the connection between right temporal and occipital region displayed significant effect in model with lag 6, 8 - 10, 13 - 20 s ($\beta$ = 0.15 - 0.18, ps < .05). The estimated marginal means for all regional connections across each model are illustrated in Fig. 4.

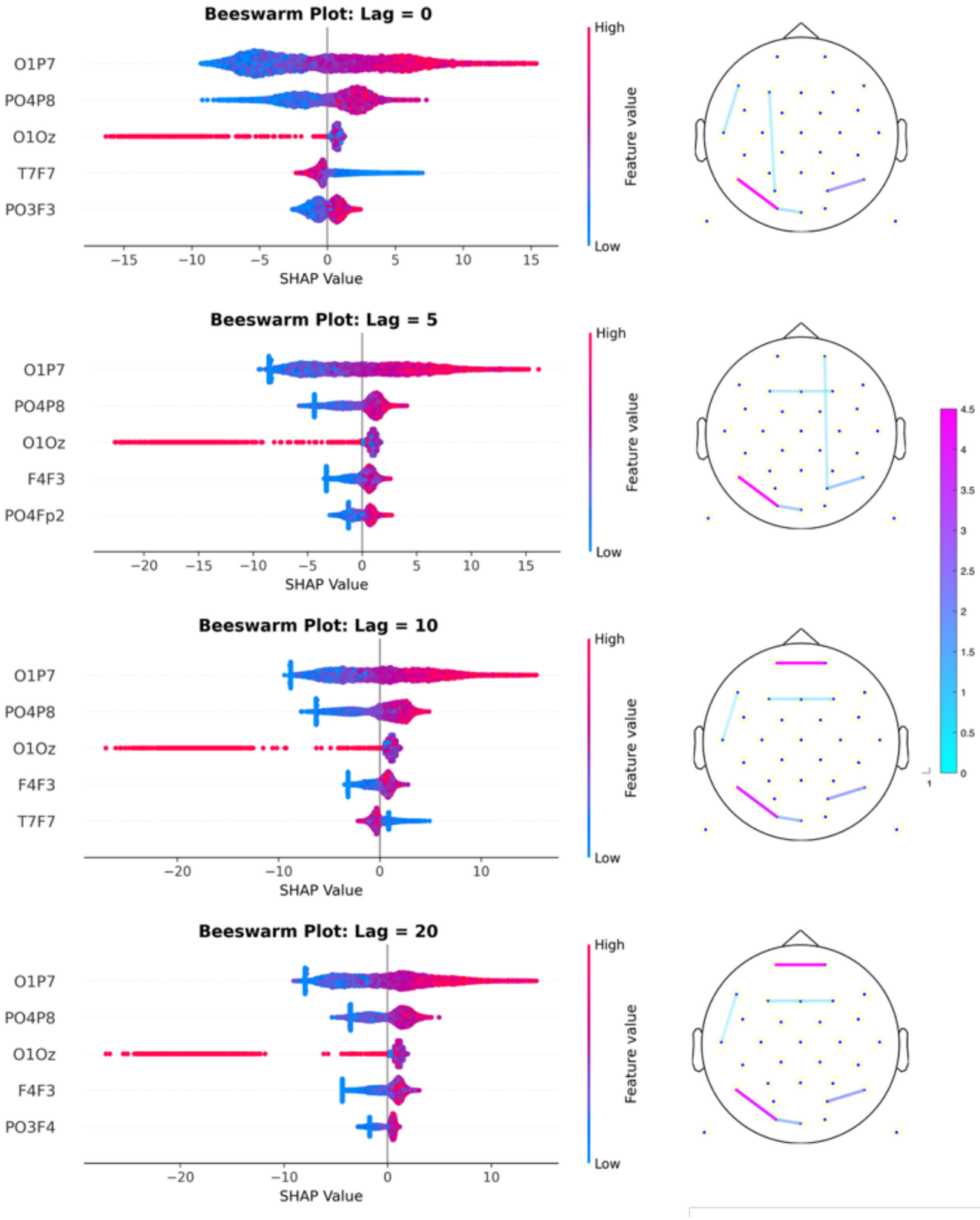

Fig. 3. **SHAP value analysis of channel-level connections for reaction time detection and forecasting models.** SHAP beeswarm plots (left) illustrate the distribution, directionality, and impact of the top five predictive features on reaction time across different time horizons (Lags 0, 5, 10, and 20 s). Corresponding topoplots (right) map the spatial distribution and relative importance of these key connections across the scalp.

## IV. DISCUSSION

The primary objective of the current study was to construct a predictive model for the forecasting of behavioral performance with high temporal resolution using EEG functional connectivity. By extracting mutual information as a measure of non-linear neural coupling feature, the random forest regression model achieved robust predictive accuracy across all time lags, demonstrating the feasibility of forecasting reaction times up to 20 seconds in advance in Psychomotor Vigilance Test (PVT). Furthermore, through the application of interpretable machine learning framework (SHAP), we disentangled the specific feature contribution underlying these predictions, identifying distinct functional connectivity biomarkers that index both the current performance impairment and the latent propensity for future performance lapses.

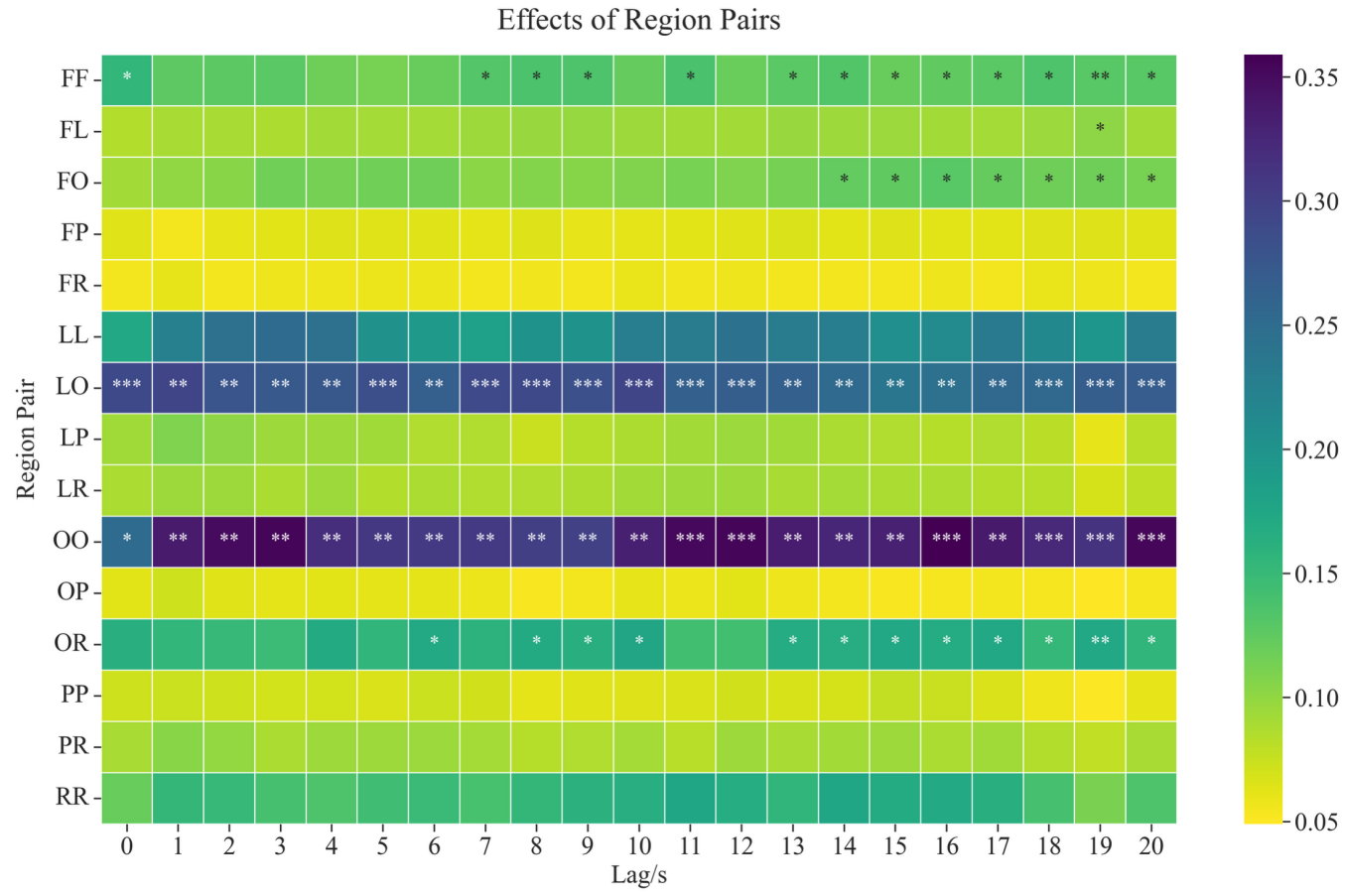

Fig. 4. **Estimated marginal means of regional connections across varying time lags**. The heatmap illustrates the predictive effects of different region pairs derived from the LME models across lags 0 to 20 s. Color intensity represents the value of the estimated marginal mean, with asterisks denoting statistical significance (*p < .05, **p < .01, ***p < .001). Region pairs are abbreviated by combining the designated letters for each region (F: Frontal, L: Left Temporal, R: Right Temporal, O: Occipital, P: Parietal; e.g., FF indicates a frontal-frontal connection, FL indicates a frontal-left temporal connection).

### *A. Fatigue Induction*

The behavioral and subjective data confirm the successful induction of mental fatigue within our experimental paradigm. Consistent with established literature [35], [39], participants exhibited a marked decline in vigilance, manifested as a significant increase in reaction time (RT) from 356 ± 24 ms to 398 ± 27 ms. This behavioral deterioration was corroborated by significant increases in subjective fatigue ratings: SSS scores increased from 2.37 ± 0.81 to 4.40 ± 1.13, and CFS scores increased from 19.47 ± 5.08 to 40.13 ± 10.80. Pearson's correlations between behavioral performance and subjective ratings further validate the successful induction of fatigue. Despite these significant correlations, a critical methodological decision in this study was to target the prediction of objective RT rather than subjective mental fatigue levels. This choice is grounded in two primary considerations. First, RT is a direct metric of behavioral impairment; in safety-critical scenarios, it is the behavioral failure (e.g., delayed response), rather than the subjective feeling of tiredness, that precipitates accidents. Additionally, subjective mental fatigue is an abstract, holistic construct that often dissociates from objective performance. Our results revealed only moderate correlations between post-task RT and subjective fatigue ratings ($r = 0.52$ and 0.38 for the SSS and CFS, respectively), supporting the phenomenon of perceptual blindness to fatigue, a dangerous state where individuals maintain a false sense of alertness despite severe cognitive degradation [40]. Second, continuous subjective monitoring is often impractical. Requiring operators to frequently report their mental state disrupts the primary task and introduces additional cognitive load, which can be hazardous in real-world settings [41]. Consequently, forecasting fatigue based on RT offers a more reliable, non-intrusive, and practical approach.

### *B. Fatigue Forecasting*

Our baseline model achieved a prediction correlation of 0.691 ± 0.014 and RMSE of 23.75 ± 0.65 ms, demonstrating superior predictive accuracy compared to recent benchmarks [12], [13], [15], thus validates the efficacy of the proposed framework. Despite the slight decline, the 20 forecasting models have sustained robustness with highest RMSE of 24.187 ± 0.570 ms and lowest $r = 0.678 \pm 0.012$, which indicates that EEG functional connectivity features contain stable, predictive information regarding future behavioral states.

Achieving a low forecasting RMSE holds substantial practical significance for real-world applications. When continuous predicted reaction times are converted into binary intervention decisions (e.g., whether to trigger an alert), higher prediction accuracy reflected by a lower RMSE directly translates to a reduced false alarm rate. This reduction is critical for user acceptance, as high false alarm rates are known to increase mental workload and erode user trust in automated systems [42], [43]. Therefore, the proposed model effectively mitigates the potential risk of alarm fatigue, and further enhances the overall operational efficacy of the fatigue forecasting system.

### *C. Model Explanation*

Analysis of feature importance calculation further elucidates the underlying neural pattern, highlighting the key biomarkers that drive fatigue forecasting.

Specifically, the functional connectivity between channel O1 and P7 emerged as a primary predictor across all models, demonstrating a strong positive correlation with reaction time ($r = 0.87 - 0.91$), suggesting a stronger connectivity predicted prolonged responses and increased fatigue. The consistent, significant impact of the left temporo-occipital connection ($\beta = 0.24 - 0.30$, all ps < .05) aligns with the sensor-level performance, which suggests that the sensor-level effects could be generalized to broader regional interactions within the temporo-occipital cortex. Given that temporal-occipital pathways are integral to the alerting and orienting networks [44], the consistent engagement of this circuit likely reflects the continuous demand for orienting attention toward visual stimuli. Its significance across all forecasting lags indicates that the maintenance of this sensory-motor interface is a critical determinant of instant performance and a robust indicator of impending cognitive failure.

Similarly, the functional connectivity between channel O1 and Oz emerged as an important feature across all models, exhibiting a strong negative correlation with reaction time (r = -0.86 to -0.81). The stable, significant impact of intra-occipital connectivity ($\beta$ = 0.25 - 0.36, all ps < .05) corresponds with the observed sensor-level patterns, which hint its negative impact, suggesting that stronger local coupling in the occipital cortex is indicative of a state of alertness and facilitates faster responses. Complementing previous findings that decreased intra-occipital connectivity is associated with vigilance decline [45], our results indicates that this local coupling serves as a robust biomarker for both fatigue detection and forecasting. Physiologically, this reduction in occipital coupling likely reflects the cortical gating of sensory stimuli, a phenomenon where the brain progressively dampens the transmission of visual inputs during drowsiness, paralleling the fading of consciousness [46]. Given that the occipital cortex is the primary hub for visual processing [47], the stability of this feature across all time lags suggests that the deterioration of visual sensory integration is a core mechanism of failure in the Psychomotor Vigilance Test (PVT).

The functional connectivity between channel F3 and F4 emerged as a prominent feature across all models, exhibiting a positive correlation with reaction time (r = 0.71 - 0.78). However, the broader regional impact of intra-frontal connectivity exhibit only sporadic significance, predominantly concentrating at longer forecasting lags (> 7 s; $\beta$ = 0.12 - 0.16, ps < .05). This temporal divergence suggests that while the impact of the specific F3-F4 connection in short-term prediction may be transient, its stable significance in longer-range forecasting reflects a broader intra-frontal dynamic. Specifically, this extended-lag coupling appears to index the brain's compensatory struggle against impending fatigue. Previous research has established the role of intra-frontal connectivity in attention sustainability [11] and executive control [48]. Our finding that these connections strengthen prior to performance decrements aligns with observations by Nguyen et al [41], suggesting a recruitment of top-down executive resources to counteract declining vigilance.

Similarly, the functional connectivity between channel Fp2 and PO4, as well as between channel F4 and PO3, demonstrated high predictive importance predominantly at longer forecasting lags (> 11 s). This temporal profile corresponds with the broader regional significance of fronto-occipital connectivity observed at lags of 14 to 20 s ($\beta$ = 0.11 - 0.13, ps < .05), indicating a positive relationship with reaction time. The enhanced frontal-occipital coupling observed in these forecasting model echoes the characteristic path length increases noted in graph-theoretic studies [39], [49], and the altered prefrontal-posterior coupling described by Gui et al [50].

Critically, we interpret these surges in connectivity as a neural signature of compensatory effort, aligning with established findings that the fatigued brain must recruit additional neural resources to maintain performance [51], [52], [53], [54]. Therefore, the elevated connectivity observed 10–20 s prior to a behavioral lapse likely reflects the increased cognitive workload required to sustain attention, serving as a precursor to performance decrements, the subsequent behavioral failure occurs when these compensatory mechanisms are ultimately overwhelmed. This temporal distinction indicated a potential dual-pattern: effective fatigue forecasting relies on detecting the brain's proactive struggle to maintain alertness, whereas immediate fatigue detection relies on detecting the effort and break down of sensory processing.

### D. Limitation

While the current study provides robust evidence for the feasibility of early fatigue prediction, several avenues for future research warrant consideration. First, although the Psychomotor Vigilance Test (PVT) is the gold standard for assessing sustained attention, its controlled nature is inherently simplified compared to complex real-world scenarios. As functional connectivity patterns may vary depending on the specific cognitive demands of the task and the performance metrics employed [22], future studies should validate these biomarkers in more naturalistic environments, such as simulated driving or operational monitoring. Second, while our model successfully established a high accuracy with pooled-data, further precision could be achieved by developing individualized prediction frameworks. Given that individuals exhibit varying degrees of cognitive resilience where performance declines at different rates over time [55], [56], [57], incorporating subject-specific calibration could further enhance model sensitivity. Finally, our study prioritized model interpretability via broadband mutual information and RF. Future investigations could complement this approach by exploring frequency-resolved connectivity features [35] or employing deep learning architectures [9] in previous detection task to capture more nuanced, high-dimensional neural representations, potentially trading some interpretability for incremental gains in predictive accuracy.

## V. CONCLUSION

This study proposed a novel model for forecasting of single-trial reaction time using EEG functional connectivity with short feature extraction windows. By leveraging mutual information and random forest regression, we demonstrated that reaction times can be accurately forecasted up to 20 seconds in advance, with model performance remaining robust across varying horizons. Beyond predictive accuracy, model explanation further revealed the rationale of high predict accuracy and provided critical neurophysiological insights. Critically, this study extends the scope of fatigue research from immediate detection to long range forecasting, a shift with profound implications for real-world deployment. The ability to predict significant cognitive performance decrements seconds in advance offers immense operational value in high-stakes environments, such as military operations and aviation.

## VI. Author Contributions

Bo Sun: Writing – original draft, Visualization, Experimental design, Software, Investigation, Formal analysis, Data curation

Liang Ma: Writing – review & editing, Resources, Methodology, Funding acquisition, Conceptualization, Project administration.

## Appendix

### A. Correspondence between channels and regions

| Chan | Region | Chan | Region | Chan | Region |
|---|---|---|---|---|---|
| Fp1 | F | FC6 | RT | Pz | P |
| Fp2 | F | Cz | P | P3 | P |
| Fz | F | C3 | P | P4 | P |
| F3 | F | C4 | P | P7 | LT |
| F4 | F | T7 | LT | P8 | RT |
| F7 | F | T8 | RT | PO3 | O |
| F8 | F | CP1 | P | PO4 | O |
| FC1 | F | CP2 | P | Oz | O |
| FC2 | F | CP5 | LT | O1 | O |
| FC5 | LT | CP6 | RT | O2 | O |

Region abbreviations: F, Frontal; LT, Left Temporal; RT, Right Temporal; P, Parietal; O, Occipital.

### B. SHAP beeswarm plots with top 5 features for all models

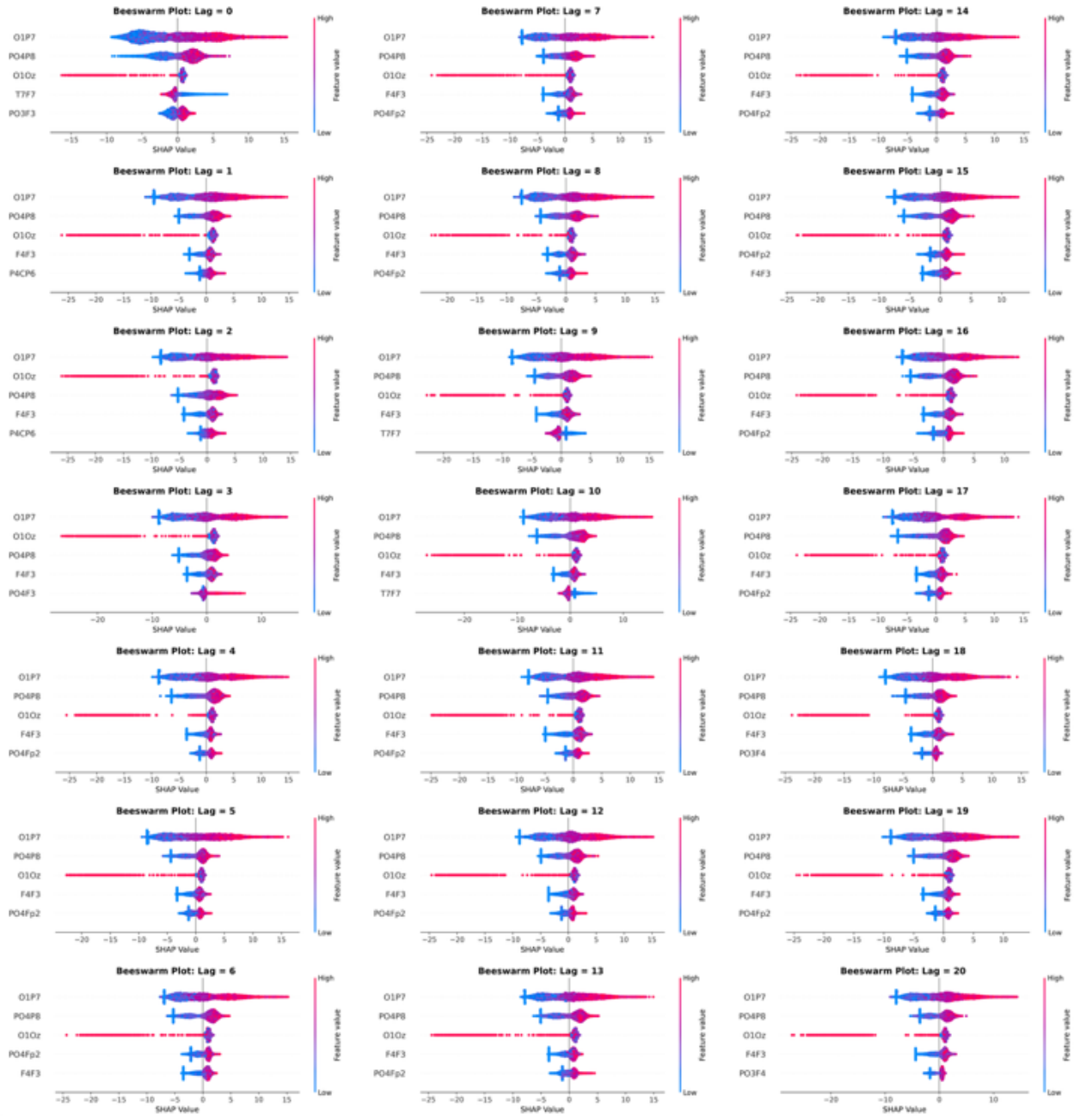

## Acknowledgment

We thank the Department of Psychology and Cognitive Science, Tsinghua University for providing the experimental environment and facilities. We thank all of the participants for their time and dedication to this study.